\documentclass[12pt, a4paper]{article}

\usepackage[utf8]{inputenc}
\usepackage{amsmath, amssymb, amsthm}
\usepackage{graphicx}
\usepackage[margin=1in]{geometry}
\usepackage{hyperref}
\usepackage{authblk} 
\usepackage{url}
\usepackage{soul}
\usepackage{rotating}
\usepackage{float}
\usepackage{lipsum}
\usepackage{subcaption}
\usepackage{xcolor}
\usepackage{apacite} 

\title{A stochastic model for fog forecasting}

\author[1]{E. Cardoso-Bihlo\thanks{Corresponding author: ecardosobihl@mun.ca}}
\author[2]{B. Khouider}

\affil[1]{Memorial University of Newfoundland, NL}
\affil[2]{University of Victoria, BC}

\date{} 

\begin{document}

\maketitle


\begin{abstract}
Despite significant advancements in parameterizations of boundary layer processes, forecasting, and nowcasting low-level clouds using numerical models remain challenging. The purpose of this study is to test a prototype of a high-resolution stochastic-deterministic model designed to simulate the life cycle of fog cover based on the Ising model from statistical mechanics. The case of advection fog around St. John's Airport in Newfoundland (Canada) has been considered.
The model demonstrates promising capabilities in forecasting mean fog cover and replicating the horizontal structure observed in satellite imagery, including bands, rolls, and closed or open cells. We evaluate the model’s predictive skill by analyzing its effectiveness in reproducing the evolution of fog cover across three representative cases. A contingency table and associated performance metrics are used to assess its accuracy.
\end{abstract}


\section{Introduction}
Predicting the occurrence, evolution, and dissipation of low-level clouds and fog remains a major challenge today \cite{roman2019radiation, bari2023fog}. In particular, fog significantly impacts human life by reducing visibility to 1 km or less, affecting aviation, sea, and road traffic, and causing damage comparable to that of hurricanes, tornadoes, and winter storms~\cite{gultepe2007fog, forthun2006trends, gultepe2019review}. 

Coastal fog along Newfoundland, Canada, is most common in July and August~\cite{fernando2021c}, making it one of the regions with the highest marine fog occurrence~\cite{dorman2017worldwide}. This fog type is challenging to forecast due to interactions between the lower atmosphere, upper ocean, and land surface processes~\cite{o2013multidecadal}. The mechanisms involving coastal fog formation are diverse. One prevalent process in the Canadian Atlantic involves warm, moist air moving over cooler coastal waters, where wind from the Gulf Stream blows over the cold Labrador Current~\cite{isaac2020characterizing}.

We propose a stochastic-deterministic fog forecasting model based on the Ising model of statistical mechanics, coupled with a vertically uniform total water content equation. To our knowledge, this is the first application of such an approach to fog. By integrating deterministic and stochastic methods, the model captures the fog life cycle. Ising-type lattice models have been successfully used in stochastic parameterization of tropical convection \cite{khouider2003coarse,khouider2010stochastic, frenkel2012using,dorrestijn2013data, khouider2014coarse, deng2015mjo, gottwald2016data, goswami2017improved,cardoso2019using}. We focus on St. John's International, one of Canada's foggiest airports, especially from April to June \cite{dorman2021large}. Despite its coastal location, sea breezes rarely form due to large-scale winds \cite{robichaud2001weather}. Local weather is shaped by wind patterns: winter winds are westerly and from southeast in the summer. Reduced visibility often coincides with northeast to southeast winds, which transport maritime air upslope due to the terrain’s elevation. 

\section{Model Configuration and Data}

The practical implementation for fog forecasting builds on theoretical studies~\cite{khouider2019new, stechmann2016cloud}. In the following, we describe the components of the model which are considered for the purpose of fog forecasting. The main components consist of the equation for the conservation of total water content following \cite{khouider2019new}:

\begin{equation} 
\frac{dq_{\text{total}}}{dt} + \mathbf{u} \cdot \nabla q_{\text{total}} = D \nabla^2 q_{\text{total}} +\frac{1}{\tau_{\text{evap}}} \Delta_{\text{surf}} q_{\text{total}} - \left(\frac{1}{\tau_{\text{turb}}} + \nabla \cdot \mathbf{u}\right) \Delta_{\text{top}} q_{\text{total}} + F
\label{eq1}
\end{equation}

The total water content  $q_{\text{total}}$ (vapour + liquid water) evolves by considering the transport of moisture through the horizontal wind $ \mathbf{u} \cdot \nabla q_{\text{total}}$, horizontal diffusion of moisture $D \nabla^2 q_{\text{total}}$, the moisture exchange at both the surface $\Delta_{\text{surf}}q$ and the top of the fog layer  $\Delta_{\text{top}}q_{\text{total}}$ through evaporation and turbulence mixing time scales $\tau _{s}$ and $\tau_{t}$ respectively and the sinks and sources of moisture  $F$ due to synoptic scale processes. In essence, F represents drying and moistening, respectively, due to large-scale subsidence and low-level moisture convergence. Here, for simplicity, this term is parameterized using in situ dew-point temperature.
\\ The lattice model consists of  $m \times m$  mesoscopic cells containing $n^2$  microscopic lattices. \\
The resulting output is the averaged Cloud Area Fraction (CAF) corresponding to the number of "foggy air" lattice sites divided by $n^2$. The moisture equation in (\ref{eq1}) is solved numerically on the mesoscopic grid and each mesoscopic grid box is associated with a $q_{\text{total}}$ value. Within each mesoscopic cell, $n^2$  lattice sites can each take on binary values (0 or 1). The Hamiltonian $H(\sigma)$ gives the total energy and dynamics of the Ising model for each configuration, where a configuration refers to a specific arrangement called spins at a specific time $t$ on the lattice:

\begin{equation}
  H(\sigma):= -\frac{1}{2} \sum_{{\bf i}, {\bf j} } J(|{\bf i}-{\bf j}|) \sigma_{\bf i}\sigma_{\bf j} + h \sum_{\bf i} \sigma_{i} 
  \label{eq2}
\end{equation}

Here $J(|{\bf i}-{\bf j}|)$ is the local interaction potential with $|{\bf i}-{\bf j}|$ being the distance between the two sites ${\bf i}$ and ${\bf j}$, in some chosen norm, and h is an external potential.
According to the theory of statistical mechanics, when the Hamiltonian system is in equilibrium, its distribution is the Gibbs canonical measure $G(\sigma)$:
\begin{equation}
    G(\sigma)=\frac{1}{Z}e^{-\beta H(\sigma)}
    \label{eq3}
\end{equation}
\vspace{0.2cm}
where $Z$ acts as normalization constant and $\beta > 0$ is the inverse temperature.

There are two coupling mechanisms between the equation for the conservation of total water content and the lattice model. The first is through $\tau_{s}$ and $\tau_{t}$ which depend on the stochastic averaged CAF and the second is through the $q_{total}$ which acts on the external potential $h$: 
\begin{equation}
    h=h_0\left(\frac{q_{total}}{q^*(T)}-1\right)
\end{equation}
where $h_{0}$ is a prefactor parameter and the ratio of the total water content $q_{total}$ to the saturation humidity $q^*(T)$, minus one, corresponds to the supersaturation inside of the mesoscale grid box  $\Delta x$ \cite{khouider2019new}.

This study integrates data from two primary sources to drive and validate a fog forecasting model. Observational ground truth for comparison is sourced from the Integrated Surface Database (ISD), a global compilation of hourly and synoptic surface observations. Specifically, data points including station identification, Date, Time, Wind Direction (Dir), Wind Speed and Gust (SPD, GUS), Present Weather odes, Temperature, Dewpoint, and Sea Level Pressure are extracted from ISD. The model's initial conditions, however, rely on ERA5 reanalysis data, and represents the fifth generation of ECMWF reanalysis. ERA5 is a comprehensive and consistent dataset of past weather and climate, derived from integrating model outputs with global observations. From ERA5, variables such as wind components, relative humidity, and temperature at 1000 hPa are extracted, all possessing an initial temporal resolution of hourly data on a 30 km grid. To accommodate predictions for the cloud area fraction, these hourly ERA5 variables are subsequently interpolated to a finer temporal resolution of 5 minutes. The wind components are interpolated as scalars using the method in~\cite{wier1995interpolating}, while other parameters are linearly interpolated. This decision is driven by the understanding that fog can form, intensify, and dissipate very quickly. However, a direct comparison of the 5-minute model output with observations is limited, as the Integrated Surface Database (ISD) only provides hourly and synoptic surface observations, preventing a direct 5-minute temporal validation against observational data. For validation, the present weather codes from the ISD are compared with the model's Mean Cloud Area Fraction (Mean CAF) to determine True Positives (TP), True Negatives (TN), False Positives (FP), and False Negatives (FN).

Equation~\ref{eq1} is solved using the Crank-Nicholson method for the diffusion and operator time-splitting with a second-order Runge-Kutta scheme for other forcing terms. The CAF is advanced on each mesoscale lattice using Gillespie’s Monte Carlo algorithm~\cite{gillespie1975exact}. Our setup uses a $496 \times 496$ km grid with 30 km resolution and a time step $\Delta t = 1$ s. The model consists of $17 \times 17$ mesoscale cells, each with microscopic sub-lattices of $n=10$. While fog cover is simulated over the full domain, the output is the averaged CAF over an area corresponding to St. John's Airport.

\subsection{Key Parameters}

The coupled stochastic-deterministic model's skill in predicting CAF, fog formation, persistence, dissipation, and $q_{total}$ depends on the key parameters. In a previous study~\cite{khouider2019new}, the model was integrated for different arbitrary values of $J$ (local interaction strength), $F$ (synoptic scale forcing), and $h_0$ (external potential strength), with fixed $\beta$ (inverse temperature). By understanding each parameter’s role, estimates can be made based on synoptic conditions conducive to fog. These parameters were verified to accurately predict fog onset and dissipation by trial and error through comparison with observations.

The strength of the local interaction potential $J$ represents the interaction between fog-free and foggy air. Given the influence of large-scale synoptic systems, advection of warm moist air over cooler waters, and local topography on fog formation at St. John's Airport~\cite{isaac2014canadian, dorman2017worldwide, dorman2021large}, we set $J = 0.2$. A positive $J$ means foggy sites make adjacent sites more likely to be foggy, mimicking small-scale processes like radiation, turbulent mixing, and latent heat release. In~\cite{khouider2019new}, $J$ values between -2 and 2 were considered.

The term $F$ represents sources and sinks of moisture, parameterizing air subsidence from the upper layer. When subsiding air moistens the fog layer, $F$ is positive; when it dries, the layer, $F$ is negative. Since F is unknown (we refer to Appendix A: Derivation of the Moisture Equation in \cite{khouider2019new}), a parameterization based on dewpoint temperature is implemented. The relative humidity (RH) and specific humidity $q$ are related by $q = RH \cdot qs$, where $qs$ is the saturation-specific humidity. In~\cite{lawrence2005relationship}, a linear RH-dewpoint relationship was identified for $RH > 50\%$. Starting with $T = T_d$, RH decreases by 5\% for every $1^\circ$ decrease in $T_d$. Based on this, we set $F$ values between $F = 30$ and $F = -250$.

The $F$ scale is constructed and tested to be experimentally accurate such that $F=30$ corresponds to  $T-T_d=0$ and decreases by a numerical value of 10 for each increment of $0.55^\circ$. 

\begin{equation}
    F = \frac{1}{\tau_F}\left[30 - \frac{340}{18.8891}(T-T_d)\right]
    \label{eq5}
\end{equation}
\vspace{0.2cm}

The parameter $h_{0}$ represents the strength of the external potential, influencing fog formation, persistence, and decay. In~\cite{khouider2019new}, higher $h_{0}$ values reduced the model's dependence on internal interactions ($J$), potentially limiting its ability to capture low-cloud interactions. Given the stronger influence of large-scale conditions in the area, we set $h_{0} = 15$, which results in faster fog expansion with some dependence on $J$ but noticeable external influence of the large-scale conditions.

Large $\beta$ values correspond to small temperature and tend toward a deterministic behavior, where the influence of $J$ and $F$ is small, while small $\beta$ values lead to highly scattered fog cover. Based on this observation, we set $\beta=5$ to reflect the intermediate modus between these two regimes. A more detailed explanation of the physical relation to fog formation and dissipation is provided in the supporting information.

\section{Case Studies: Evaluating Fog Forecasting Performance Using Mean CAF and Observed Weather Conditions}

To assess the effectiveness of the model in predicting fog occurrences, we have selected 3 separate fog events around St. John's airport. These cases represent various fog regimes and weather conditions. Our analysis focuses on comparing the duration of fog cover in terms of the predicted mean cloud fraction (CAF) with the recorded weather codes (44, 45, and 48), indicative of different fog conditions.

We use time series plots for Mean CAF, CAF skewness (SKW), and CAF standard deviation (STD). Negative SKW values indicate near-saturation fog coverage, while positive values suggest scattered fog patches with clearer areas. STD reflects variability in fog cover; higher STD means more variation, while lower STD indicates more uniform coverage. Note that the domain in the snapshots is different from the model output domain, which is used to calculate CAF, SKW, STD, and RH. The domain’s large-scale RH is calculated from specific and saturation-specific humidity. 

\begin{figure*}[t!]
    \centering
    \begin{minipage}[t]{0.48\textwidth} 
        \centering
        \begin{subfigure}[t]{\textwidth}
            \centering
            \includegraphics[height=1.5in]{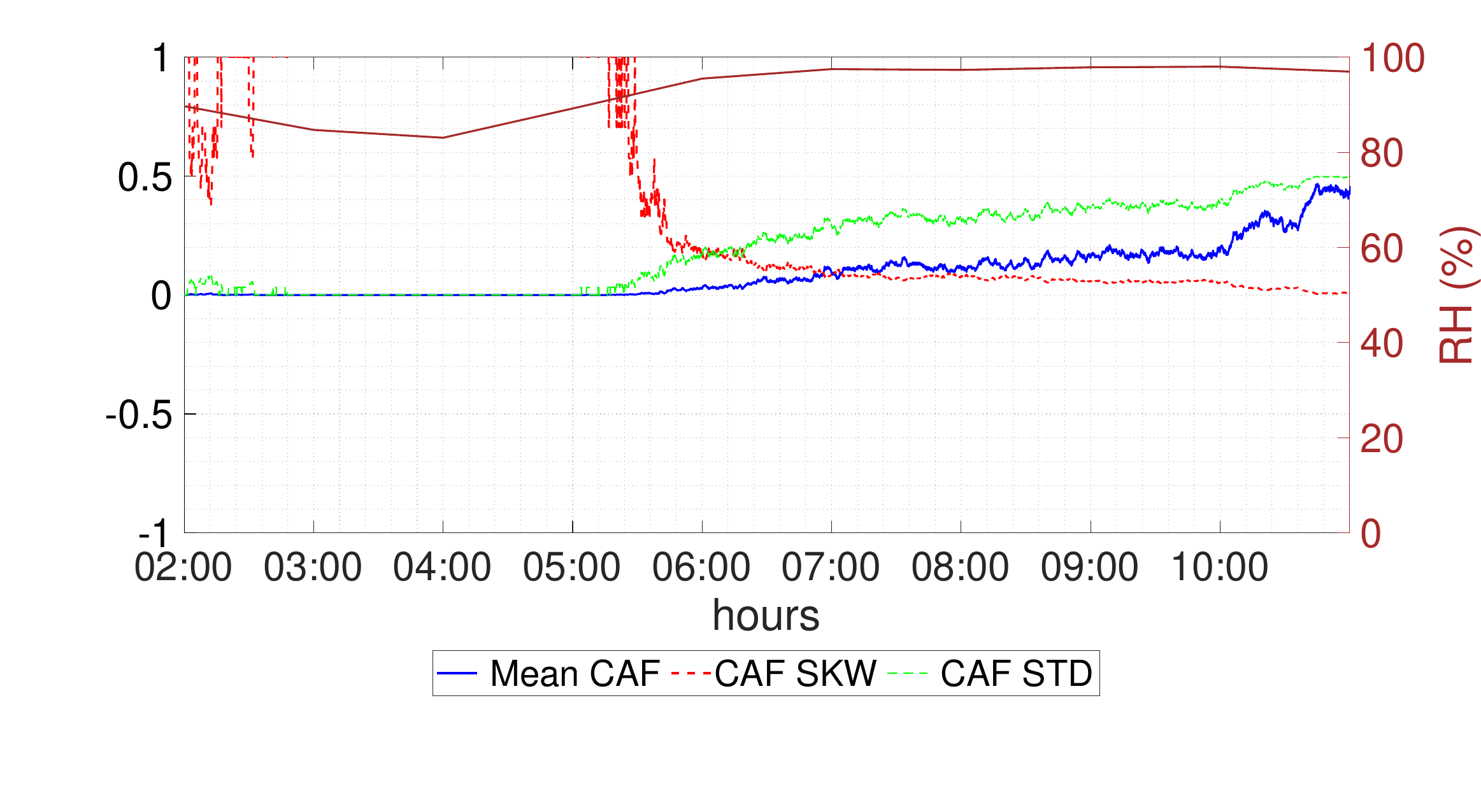} 
            \caption{Model: May 5, 2014, time: 2:00 -11:00 }
            \label{fig:img1}
        \end{subfigure}
        \vfill 
        \begin{subfigure}[t]{\textwidth}
            \centering
            \includegraphics[height=1.5in]{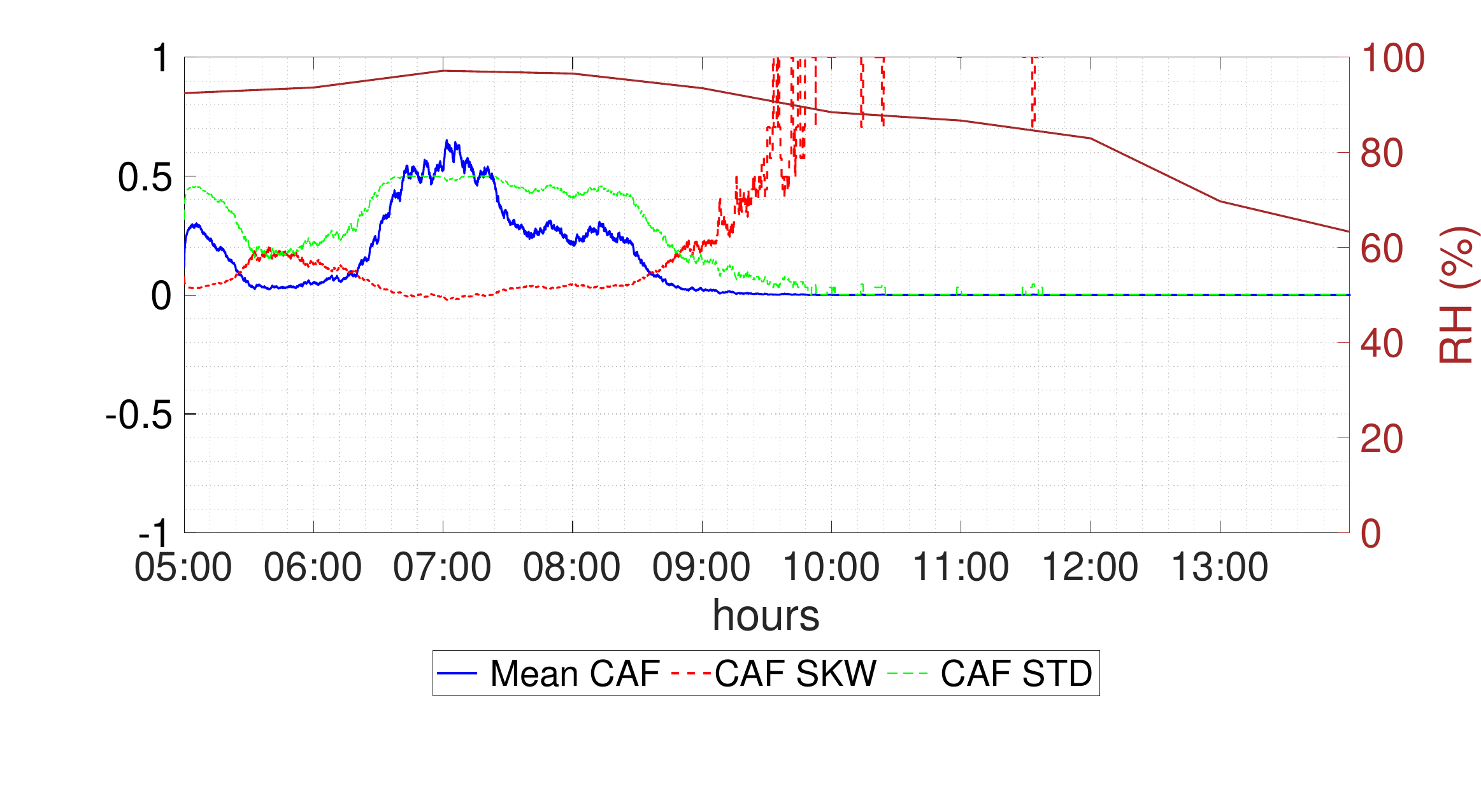} 
            \caption{Model: May 30, 2014, time: 5:00- 15:00}
            \label{fig:img2}
        \end{subfigure}
        \vfill 
        \begin{subfigure}[t]{\textwidth}
            \centering
            \includegraphics[height=1.5in]{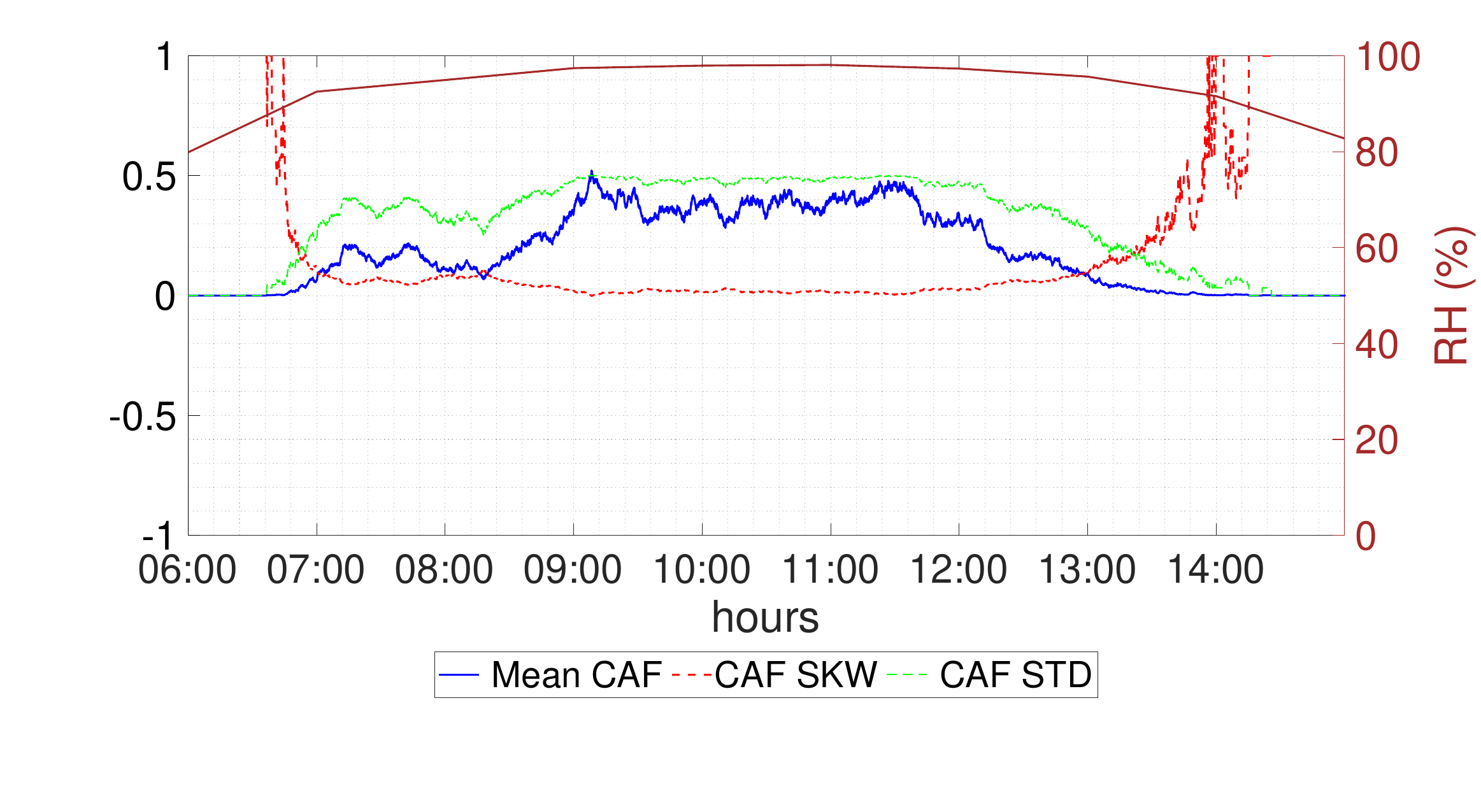} 
            \caption{Model: May 7, 2014, time: 6:00 to 15:00}
            \label{fig:img3}
        \end{subfigure}
    \end{minipage}%
    \quad 
    \begin{minipage}[t]{0.48\textwidth} 
        \centering
        \begin{subfigure}[t]{\textwidth}
            \centering
            \includegraphics[height=1.3in]{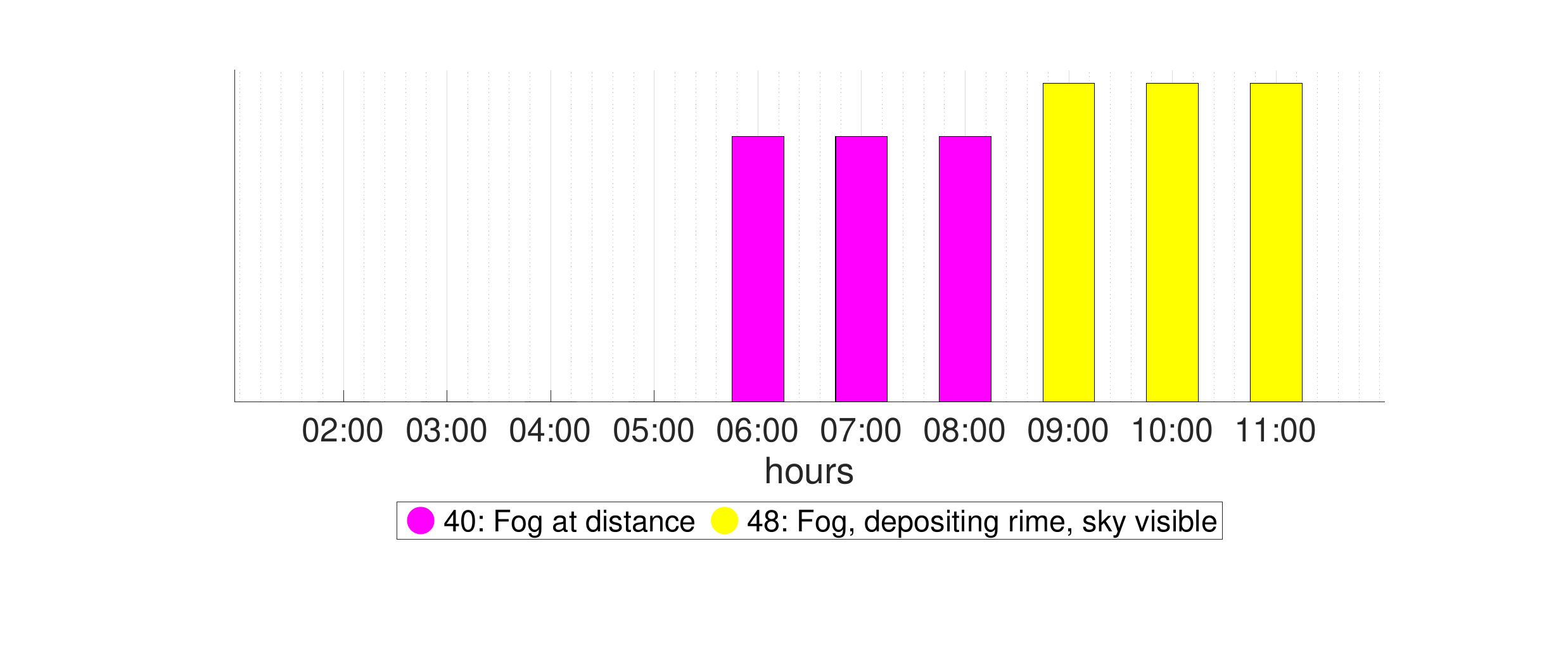} 
            \caption{Observations:  May 5, 2014, time: 2:00-11:00 }
            \label{fig:img4}
        \end{subfigure}
        \vfill 
        \begin{subfigure}[t]{\textwidth}
            \centering
            \includegraphics[height=1.3in]{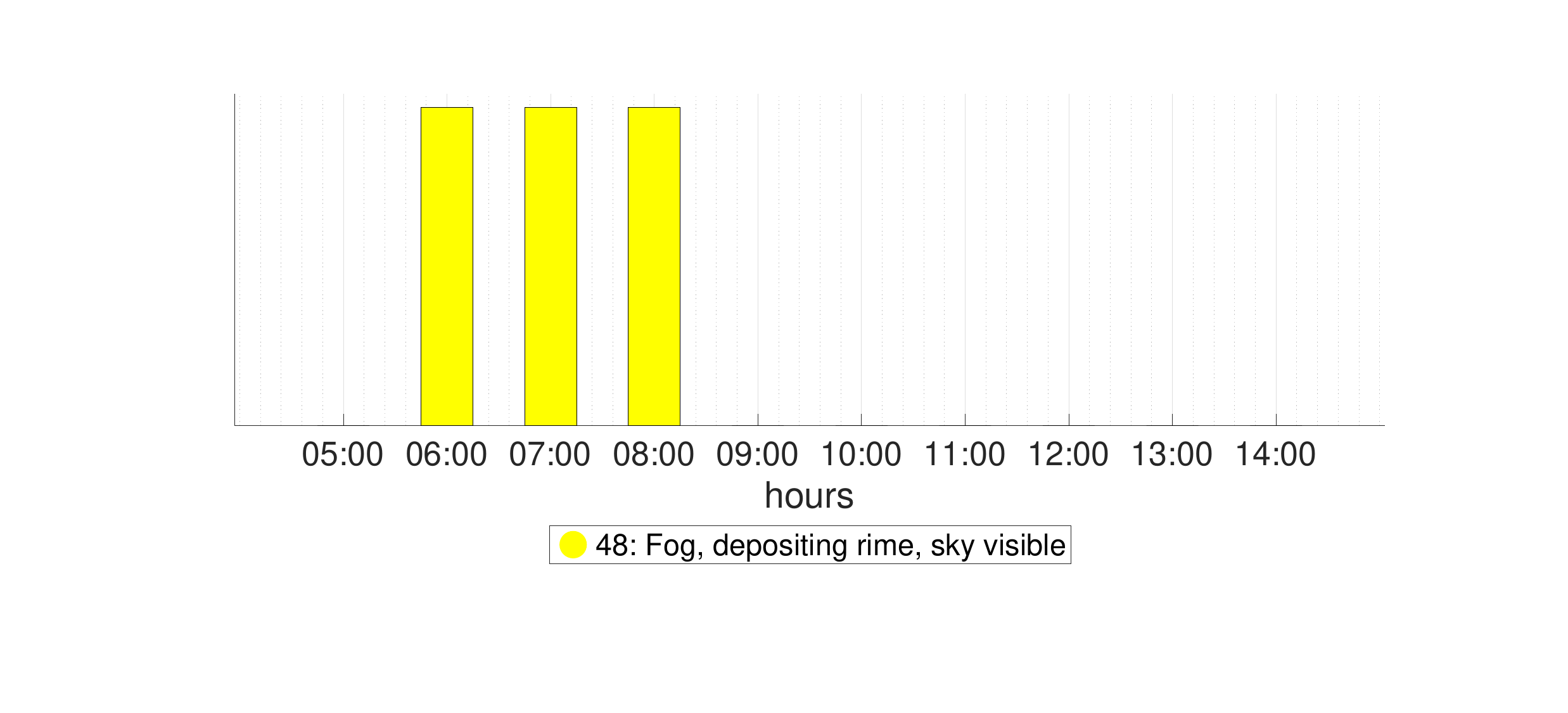}  
            \caption{Observations: May 30, 2014, time: 5:00 to 15:00}
            \label{fig:img5}
        \end{subfigure}
        \vfill 
        \begin{subfigure}[t]{\textwidth}
            \centering
            \includegraphics[height=1.3in]{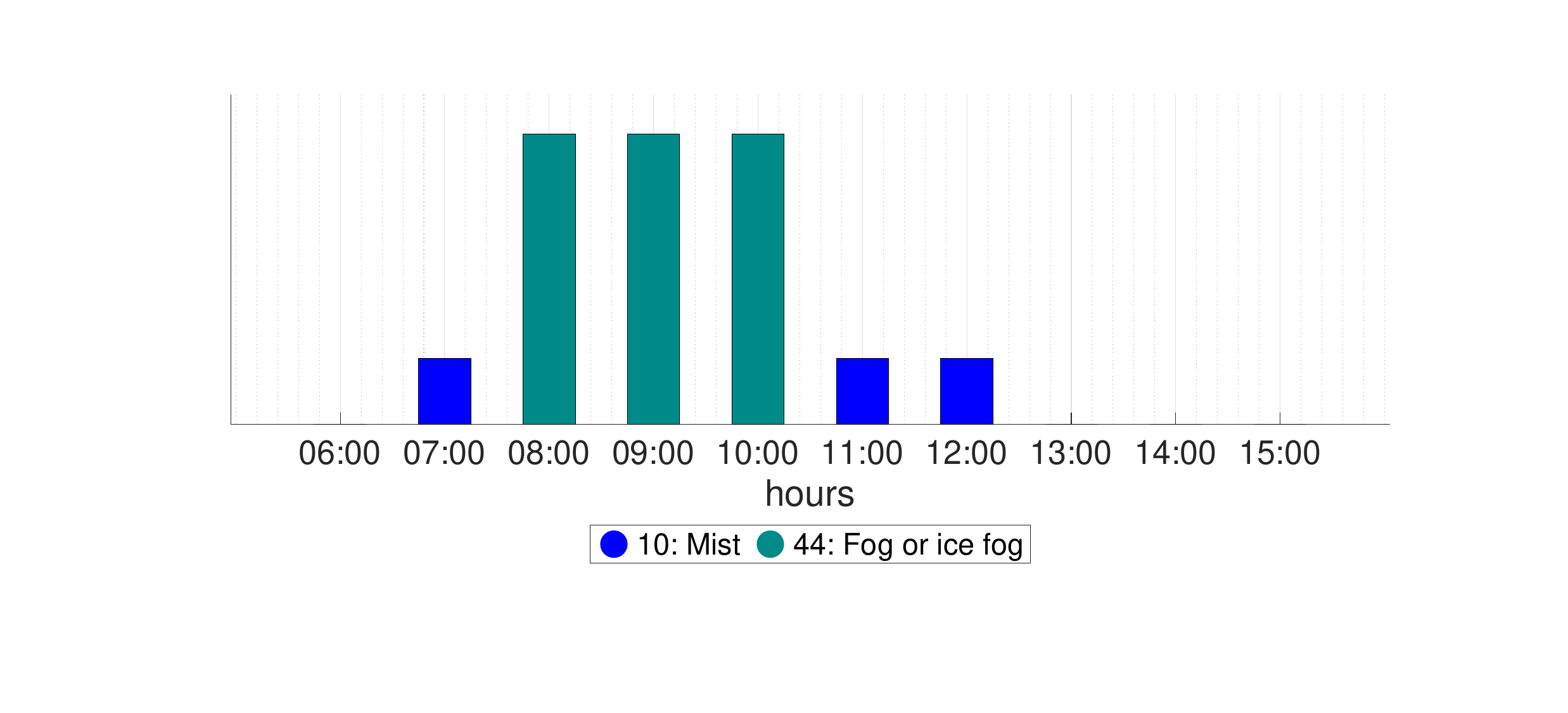}  
            \caption{Observations:  May 7, 2014, time: 6:00 to 15:00}
            \label{fig:img6}
        \end{subfigure}
    \end{minipage}
    \caption{Time series of modeled Mean CAF, CAF Skewness, CAF Standard Deviation, and Relative Humidity (RH). Right: Corresponding weather observations at St. John's Airport, where different fog types are distinguished by a unique color and bar height.}
    \label{ }
\end{figure*}

\begin{figure}[htb]
    \begin{minipage}[t]{.32\textwidth}
        \centering
        \includegraphics[height=3.5cm, keepaspectratio]{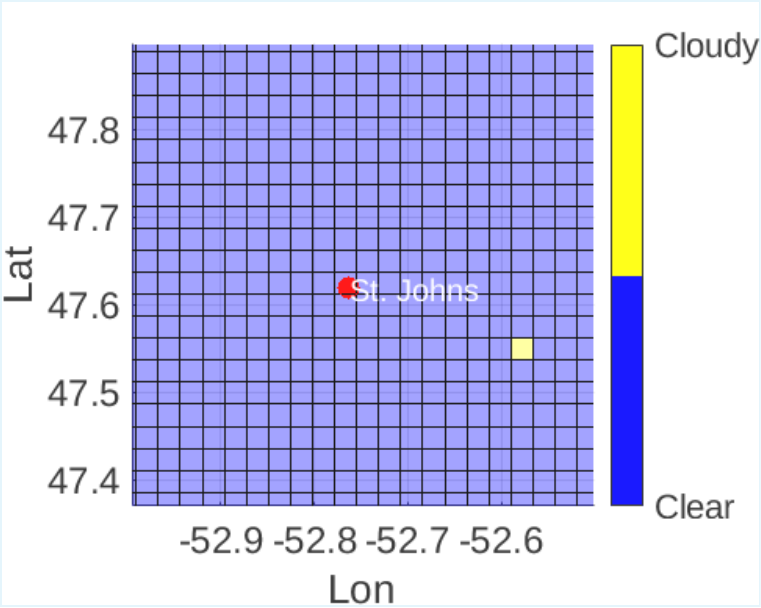}
        \subcaption{2:05 am, May 5 2014 }\label{fig:img7}
    \end{minipage}
    \hfill
    \begin{minipage}[t]{.32\textwidth}
        \centering
        \includegraphics[height=3.5cm, keepaspectratio]{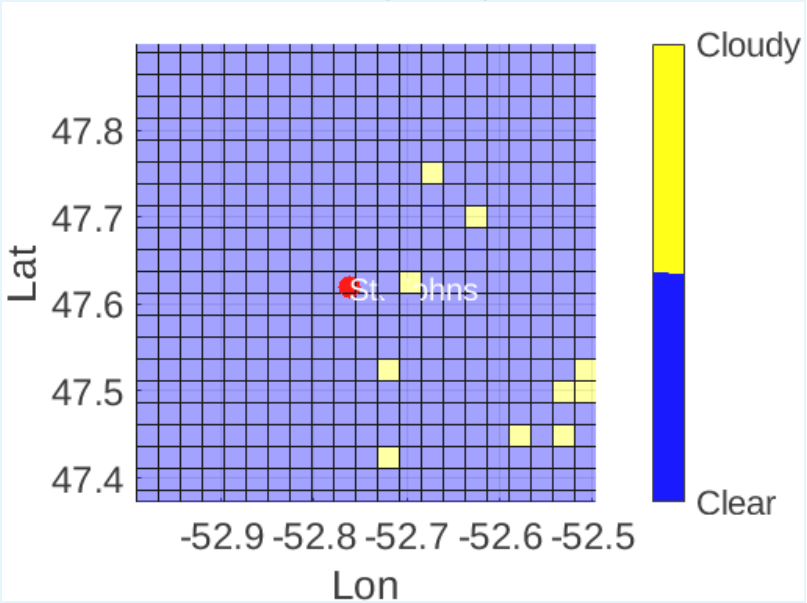}
        \subcaption{ 6:00 am, May 5 2014 }\label{fig:img8}
    \end{minipage}
    \hfill
    \begin{minipage}[t]{.32\textwidth}
        \centering
        \includegraphics[height=3.5cm, keepaspectratio]{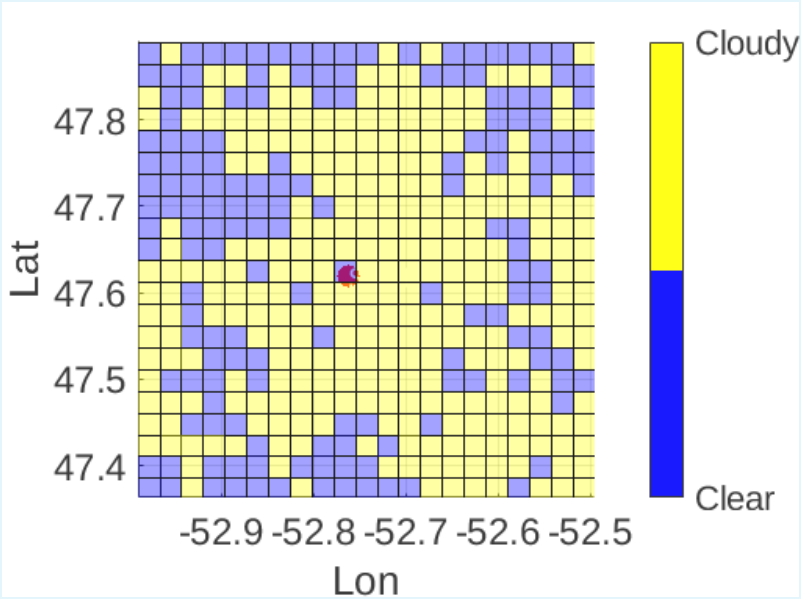}
        \subcaption{11:00 am, May 5 2014}\label{fig:img9}
    \end{minipage}
    
    \begin{minipage}[t]{.32\textwidth}
        \centering
        \includegraphics[height=3.5cm, keepaspectratio]{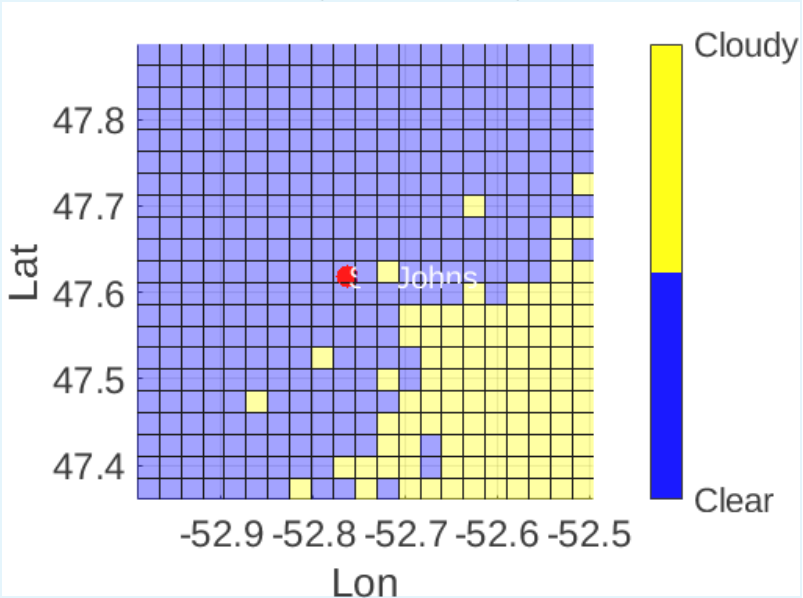}
        \subcaption{5:05 am, May 30,2014}\label{fig:img10}
    \end{minipage}
    \hfill
    \begin{minipage}[t]{.32\textwidth}
        \centering
        \includegraphics[height=3.5cm, keepaspectratio]{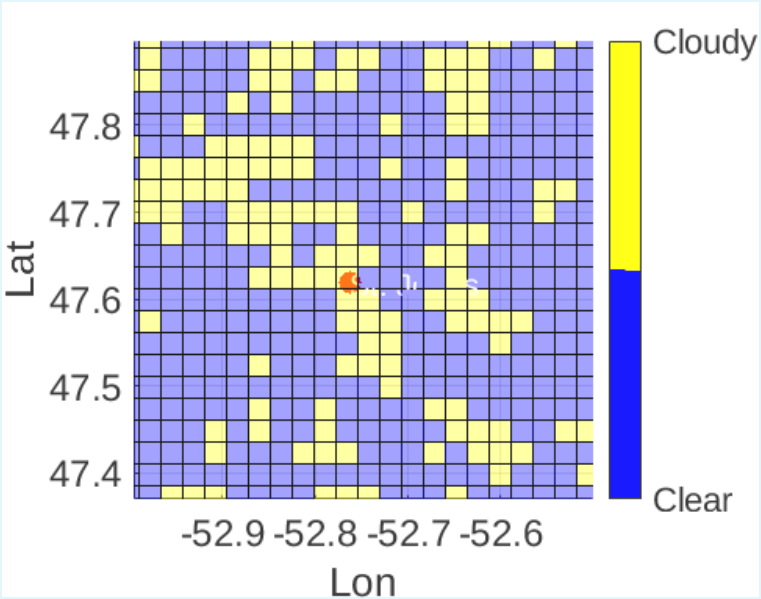}
        \subcaption{6:00 am, May 30, 2014}\label{fig:img11}
    \end{minipage}
    \hfill
    \begin{minipage}[t]{.32\textwidth}
        \centering
        \includegraphics[height=3.5cm, keepaspectratio]{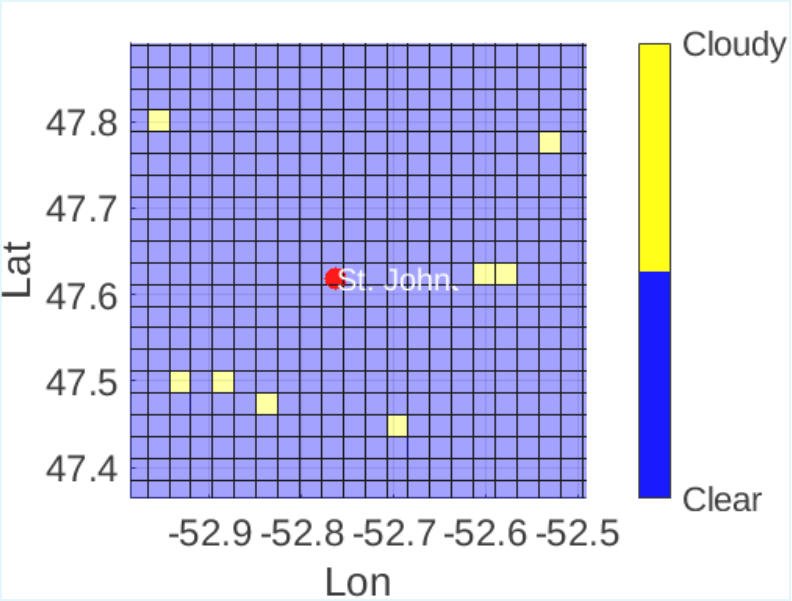}
        \subcaption{9:00 am, May 30, 2014}\label{fig:img12}
    \end{minipage}
    
    \begin{minipage}[t]{.32\textwidth}
        \centering
        \includegraphics[height=3.5cm, keepaspectratio]{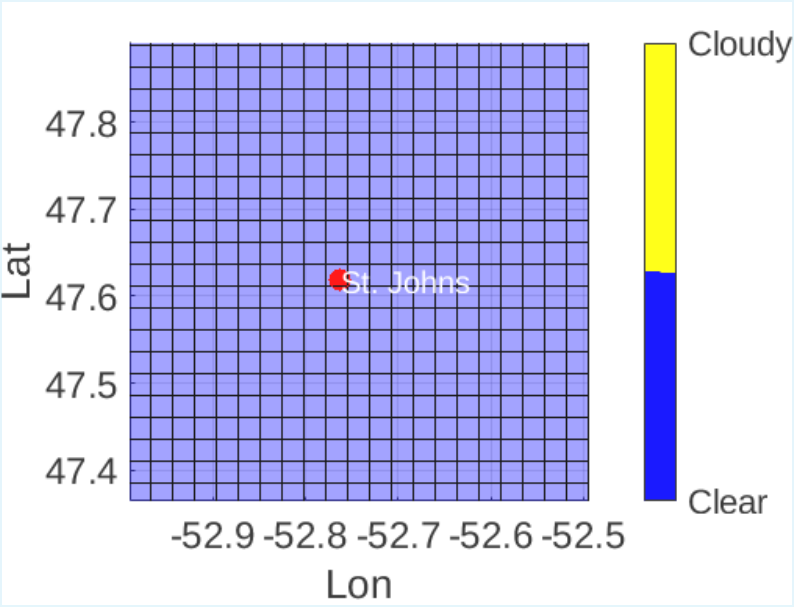}
        \subcaption{6:40, May 7 2014 }\label{fig:img13}
    \end{minipage}
    \hfill
    \begin{minipage}[t]{.32\textwidth}
        \centering
        \includegraphics[height=3.5cm, keepaspectratio]{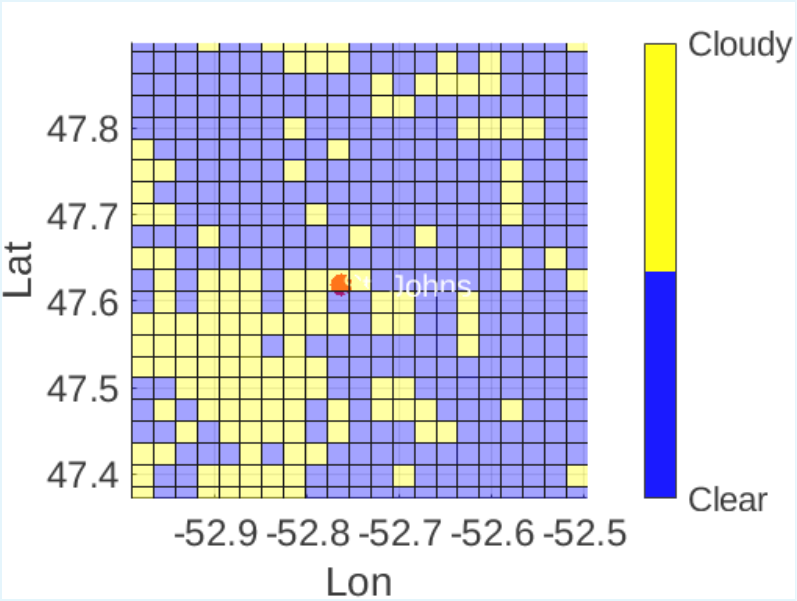}
        \subcaption{10:10, May 7 2014 }\label{fig:img14}
    \end{minipage}
    \hfill
    \begin{minipage}[t]{.32\textwidth}
        \centering
        \includegraphics[height=3.5cm, keepaspectratio]{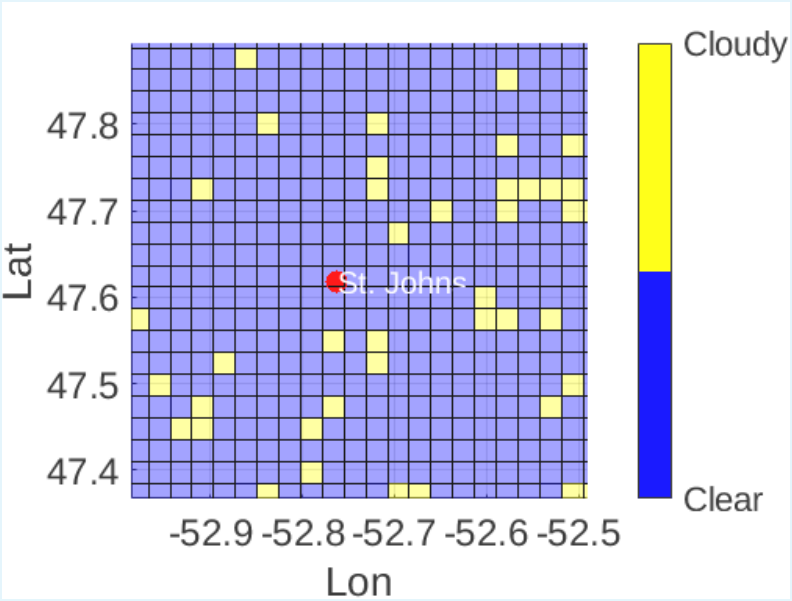}
        \subcaption{13:00,  May 7 2014}\label{fig:img15}
    \end{minipage}  
    
    \caption{Snapshot of the binary $\sigma_i$ for the simulation on May 5, 2014 (a, b,c), May 30, 2014 (d,e,f), and May 7, 2016 (g,h,i).}\label{fig:all-nine-images}
\end{figure}

\subsection{May 05 2014, time 2:00-11:00}

On May 5th, 2014, winds shifted from south to southeast, with speeds ranging from 2-7 m/s, increasing to 11 m/s with gusts. Temperature remained around $0-1^\circ$ from 2:00-9:00, with a dew point difference of 1°C. Sea-level pressure increased from 1016 hPa to 1020 hPa (not shown). Fog formed, reducing visibility to 1 km at St. John's airport 
by 9:00. Figure~\ref{fig:img4} shows weather codes: '40' (fog at a distance) from 6:00 for three hours, followed by '48' (fog, depositing rime). In Figure~\ref{fig:img1} it can be observed that the model effectively simulates the time of initiation of the fog cover, as evidenced by a Mean CAF value of 0 in the first 4 hours of the simulation, followed by a slight increase in CAF and decrease skewness to zero in the subsequent hour which marks the recorded fog observation at the St. John's airport. In Figure~\ref{fig:img8}, snapshots at 6:00 show fog cells, which coincide with the recorded weather code '40'. By 11:00 (Figure~\ref{fig:img9}), fog coverage increased, matched by the rising Mean CAF. The RH values are also observed to increase to nearly 100\% matching fog observation over St. John's (Figure~\ref{fig:img1})

\subsection{May 30 2014, time 5:00-15:00}

Observations indicate that the wind initially blew from the east, shifted to the northeast during the fog event, and returned eastward. Wind speeds remained gentle (0--2 m/s) during the fog event but increased to 5--7 m/s afterward. The dew point matched the temperature throughout, with temperatures rising from $-2^\circ$ to $0^\circ$ after fog dissipation. Atmospheric pressure fluctuated around 1017 hPa during the fog event, then increased to 1021.3 hPa (not shown). 
Figure~\ref{fig:img5} depicts a short fog event occurring between 6:00 and 8:00. At 5:00, the model output indicates a positive mean CAF as shown in Figure~\ref{fig:img2}, but no fog covered St. John’s station (Figure~\ref{fig:img5}). Fog cells extended inland from the southeast, as is typical for early mornings in St. John’s (Figure~\ref{fig:img10}). By 6:00 (Figure~\ref{fig:img11}), fog was present at the St. John's station. Dissipation became prominent between 8:00 and 9:00, with Mean CAF dropping to zero and fog vanishing by late morning. Relative humidity exceeded 90\% as fog formed and declined as it dissipated (Figure~\ref{fig:img2}).

\subsection{May 7, 2016, time 6:00-13:00}

On May 7th, 2016, during the first hour of observation, winds from the west reached 17 m/s with occasional gusts. By the second hour, winds shifted northwest, and fog began to form as wind speed decreased. The cloud ceiling dropped to 1 km, with visibility reducing to 1 km in the following hour. Afternoon, visibility improved as the fog dissipated. The temperature remained around 7°C, with the dew point staying within 0-0.5°C of the temperature. Sea-level pressure (SLP) rose from 1011.9 hPa in the morning to 1017.4 hPa by mid-afternoon (not shown). As displayed in Figure~\ref{fig:img6}, mist (blue bar) was observed at 7:00 (weather code 10), fog occurred from 8:00 to 10:00, and mist reappeared at 11:00, lasting until early afternoon. The CAF showed a slight increase during the second hour of the simulation, as seen in Figure~\ref{fig:img3}, while mist was recorded. Fog cover steadily increased after 10:10, as shown in the snapshot of Figure~\ref{fig:img14}, corresponding with the mist event. At noon, CAF began to decrease, as shown in Figure~\ref{fig:img3}. The initial snapshot at 6:20 in Figure~\ref{fig:img13} showed no cells around St. John’s airport, but by 10:10, elongated cells developed over and around the airport. By 13:00, the fog cells spread and reduced in number, aligning with the dissipation of fog observed in Figure~\ref{fig:img15}.

\section{Model validation}

Beyond the three case studies, we ran the model 95 times for independent 10-hour samples from May 2014–2017 for verification purposes. The mean CAF was compared with observations (CAF $>$ 0: fog, CAF = 0: no fog), where 40\% of the total number of hours corresponds to fog events. In categorical verification, we compare binary forecasts and observations using a contingency table: true positives: 354, false positives: 98, false negatives: 35 and true negatives: 463. The resulting performance metrics are Accuracy: 86\%, Precision: 78\%, Recall: 91\%, False Positive Rate: 21.6\% and F1 Score: 84\%.

\section{Summary}

To forecast fog formation and dissipation, meteorologists use satellite, radar, models, and climatology.  In this study, we present a model that integrates both deterministic and stochastic processes, tested on three advection fog events at St. John's Airport. 
The case of May 5, 2014, illustrates a progression from "Fog at distance" (Code 40) to "Fog, depositing rime, sky visible" (Code 48), which is observable in the binary sigma plots for that day. This particular instance can be classified as a case demonstrating the initiation of fog over the ocean and its subsequent expansion inland.
The May 30, 2014, case presents a very brief episode of fog (Code 48), commencing in the early morning and dissipating within two hours. This event was chosen to show the model's potential skill in reproducing “short” fog events.
Finally, the May 7, 2016, case displays a more complex sequence, transitioning from "Mist" (Code 10) to "Fog or ice fog" (Code 44), and then reverting to "Mist" (Code 10). This specific example demonstrates a complete fog lifecycle, encompassing its initiation, persistence, and eventual dissipation, which is also well reproduced by the model.
The model predicts fog development accurately, correlating with observed visibility and weather codes. Model snapshots reveal fog cell distribution and evolution. Fog formation is particularly frequent in coastal and inland areas during the morning hours, especially in parts of the Avalon Peninsula, where St. John's International Airport is located, as documented by the three case studies presented in this study. The fog model demonstrates strong overall performance with an accuracy of 86\%, correctly classifying the majority of cases. It detects fog events with a high recall of 91\%, though its precision of 78\% indicates some false positives. The F1 score of 84\% suggests a good balance between precision and recall. 

\section{Open Research}

ERA5 data was downloaded from the European Center for Medium-Range Weather Forecasts (ECMWF), Copernicus Climate Change Service (C3S) at Climate DataStore (CDS) (pressure levels) \cite{hersbach2023era5}.
The datasets from Integrated Surface Database (ISD) were accessed through \cite{smith2011integrated} and are available from the NOAA National Centers for Environmental Information (NCEI) platform at \url{https://www.ncei.noaa.gov/products/land-based-station/integrated-surface-database}.

\section*{Acknowledgments}
We would like to thank Dr. Alex Bihlo for the interesting and helpful discussions that greatly contributed to this work.\\

\noindent \textbf{Conflict of Interest}:
The authors declare no conflicts of interest relevant to this study.

\end{document}